\documentclass[11pt]{article}

\usepackage[preprint]{acl}

\usepackage{times}
\usepackage{latexsym}

\usepackage[T1]{fontenc}

\usepackage[utf8]{inputenc}

\usepackage{microtype}

\usepackage{inconsolata}

\usepackage{graphicx}
\usepackage{amsmath,amssymb}
\usepackage{booktabs}
\usepackage{multirow}

\title{How Fragile Is Safety Alignment at Frontier Scale? \\ A Single-Direction Attack on a 320B MoE}

\author{
  \textbf{Yi Shi}, \textbf{Tanyu Chen}, \textbf{Kai Shen}
\\
  Continuum AI
\\
  \texttt{research@orcarouter.ai}
}

\begin{document}
\maketitle
\begin{abstract}
Directional ablation removes an aligned language model's ability to refuse by projecting a single ``refusal direction'' out of the weights that write the residual stream. It needs no gradient-based training and no optimization, only a few hundred contrastive prompts, which makes it the canonical white-box attack on open-weight alignment. However, it has been established only on dense models up to roughly 70B parameters. We study whether it survives the shift to frontier mixture-of-experts (MoE) models whose residual streams are no longer a single tensor and whose weights ship quantized. We apply it to GLM-5.3-Flash (320B parameters, 288 routed experts, a four-wide hyper-connection residual, block-FP8). The attack survives the architecture, but what it reaches is no longer where a reader of the original recipe would look for it. Editing the attention, dense and routed-expert writers on their own removes 0.039, 0.016 and 0.148 of refusal respectively; editing all three together removes 0.776. As a result, 74\% of the effect exists only under the joint intervention. The part the conventional recipe reaches by module-name matching accounts for 0.066 of that 0.776, which is why it fails \emph{silently} on an MoE. The effect does not follow from removing just any direction: ablating a random direction orthogonal to it leaves refusal unchanged. A category-concentrated residue survives every edit we tried: subspaces fitted on violence, sexual content and hate leave measurable refusal at every rank from 1 to 12. We report the method, the 41--89 percentage-point reductions it achieves across seven harmful benchmarks with no detected change in capability, and the boundary where it stops.
\end{abstract}

\section{Introduction}

Aligning a language model to refuse harmful requests \citep{ouyang2022instructgpt, bai2022constitutional, ganguli2022redteaming, touvron2023llama2} is, mechanistically, the installation of a behavior that a white-box adversary may later try to remove. The cheapest known removal is \emph{directional ablation} \citep{arditi2024refusal}: refusal is mediated by a single linear direction in the residual stream, and orthogonalizing the weights that write that stream against the direction strips the behavior while leaving the model otherwise intact. It requires no gradient-based training, no optimization, and no access beyond the checkpoint itself. Anyone who can download an open-weight model can run it. This makes the geometry of refusal a security property of every released model rather than an academic curiosity. Yet the finding that refusal is a single direction was established on dense models up to roughly 70B parameters.

Frontier open weights have since diverged from the model this method assumes in three directions at once. Dense feed-forward blocks have become sparse mixtures of experts; the single residual stream has become the several parallel, per-layer-mixed streams of a hyper-connection residual; and the evidence that refusal is one direction, gathered on small models, has to carry to an untested frontier scale. Each divergence unsettles one of directional ablation's premises. Sparsity unsettles the question of where to cut: the conventional recipe edits the dense projections and deliberately skips the router to preserve expert routing, but recent work locates safety-sensitive behavior inside the expert representations rather than in the routing decisions themselves \citep{safetyexperts2026}, and work extending steering to MoE models finds that the refusal signal it recovers does not coincide with expert routing at all, assigning a substantial role to attention \citep{expertrefusalsteering2026}. A multi-stream residual unsettles how to cut at all, since hyper-connections \citep{zhu2025hyperconnections} and their manifold-constrained variant mHC \citep{mhc2025} carry several parallel streams mixed per layer, and a projection applied to one stream has no obvious relation to what a later layer reads. Underneath both sits the single-direction hypothesis, established on dense models up to roughly 70B parameters \citep{arditi2024refusal} and since complicated by work that finds refusal mediated by several independent directions, or by an entire ``concept cone'' \citep{wollschlaeger2025geometry, morethanone2026, refusalaffine2024}, which at frontier scale one direction may not capture. These premises matter operationally: a defender deploying open weights needs to know which part of the network the attack reaches, whether the recipes circulating online actually work at this scale, and how much of a model's safety training is robust to removal at all.

We answer these questions on GLM-5.3-Flash, which exhibits all three shifts at once: 320B parameters, 288 routed experts (sparse), a four-wide hyper-connection residual (multi-stream topology), and block-FP8 weights (quantized). We apply directional ablation and document both what the method achieves and where it stops. Our subject is the \emph{process}, not an evaluation of the model's safety: where refusal lives in such a model, how the architecture and quantization must be handled, and the boundary the method reaches. The work is also instrumental: security research needs models that will engage with adversarial material, from malware behavior to exploitation logic to phishing patterns, and a model that refuses on surface features cannot support that analysis. Understanding what it takes to remove refusal, and what refuses to be removed, is therefore a prerequisite to building security tooling on open weights.

The premises do not fail alike, and the first to go is the assumption that refusal is written somewhere in particular. It has no answer at the level of a single group of weights: editing attention, the dense and shared projections, or the routed experts on their own removes 0.039, 0.016 and 0.148 of refusal, while editing all three together removes 0.776. Three quarters of the effect is not attributable to any group alone. What the conventional recipe reaches by module-name matching---attention and the dense projections---removes 0.066 of that 0.776, because the fused expert parameters are invisible to the module traversal it relies on; it reports a clean edit and changes almost nothing, which is what makes the failure silent. The multi-stream residual, by contrast, does not obstruct the attack: its mixing weights are scalar, so a direction that is zero in every stream stays zero, and editing the residual-writing weights turns out to be more thorough than hooking the layer output. The attack does not work by removing just any direction: ablating a random direction orthogonal to it leaves refusal where it was. Yet a fraction of refusal, concentrated in a few content categories, survives every subspace we fitted, up to rank 12. Across seven harmful benchmarks the method removes 41--89 percentage points of refusal with no detected change in capability. The width of that range is itself the boundary, since the benchmarks it clears least are the ones dense in the resistant categories.

\noindent Our contributions are: \begin{itemize} \item We decompose the edit's effect over all seven subsets of the writer groups in a 320B mixture-of-experts and find it strongly non-additive: no group removes much alone, and three quarters of the joint effect appears only when attention, the dense projections and the routed experts are edited together. The subset a name-matching implementation reaches removes 0.066 of an available 0.776, which is the silent-failure mode the conventional recipe hits on an MoE.
\item We show that directional ablation is compatible with a hyper-connection residual, and that editing the residual-writing weights is more thorough than hooking the layer boundary, a difference detected by a paired comparison on the same prompts.
\item We empirically characterize the reach and limitations of the tested low-rank edits. The effect does not follow from removing an arbitrary direction: a random orthogonal direction removes nothing. But a category-concentrated residue survives subspaces fitted specifically on those categories at every rank from 1 to 12, and we report which categories they are.
\end{itemize}

Because the attack is already public and the checkpoint we study is uncensored, we defer threat model and release conditions to \S\ref{sec:ethics}. There we argue that documenting where safety training is and is not robust to a known attack helps defenders more than it helps attackers.

\section{Background}

That an aligned language model refuses harmful requests is a behavior acquired through safety training, and interpretability work has shown that this behavior is represented by low-dimensional linear structure in the residual stream \citep{arditi2024refusal, wollschlaeger2025geometry}. Directional ablation exploits exactly this: estimate one direction from a few hundred contrastive prompts, project it out of the weights that write the residual stream, and the behavior is gone, with no gradient steps and no optimization. The validity of that construction, however, rests on premises about how a model is built: that the residual stream is a single tensor, that the weights writing to it can be enumerated exhaustively, and that refusal is mediated by one linear direction within it. On the generation of models where the method was established, dense Transformers of at most roughly 70B parameters, these premises hold so naturally that they are rarely stated. Frontier open weights have since moved as a whole. Sparse mixtures of experts have replaced dense feed-forward blocks as the dominant route to scaling parameter count \citep{jiang2024mixtral, deepseekv3}. The residual stream has begun to be replaced by several parallel streams mixed layer by layer \citep{zhu2025hyperconnections, mhc2025}. Whether the premises still hold is no longer something that can be assumed.

\subsection{Directional Ablation}

Write the residual stream at layer $l$ as $h^l \in \mathbb{R}^d$. \citet{arditi2024refusal} take the refusal direction to be the difference in means between harmful and harmless prompts over the last-token representation at some layer, $r = \mathbb{E}_{\text{harmful}}[h^l_{-1}] - \mathbb{E}_{\text{harmless}}[h^l_{-1}]$, normalized to unit length. To remove refusal, every weight matrix $W$ that writes to the residual stream---the embedding matrix $W_E$ together with the attention output projections $W_O^l$ and the feed-forward down-projections $W_{\text{down}}^l$, which are where a feed-forward block writes its output back \citep{geva2021ffn}, at each layer---is replaced by $W \leftarrow W - r r^{\top} W$, which projects each writer's output onto the orthogonal complement of $r$; thereafter the direction $r$ is never written into the stream. The operation needs no gradients and no optimization, only the prompts the two means are taken over. On dense open-weight models of up to roughly 70B parameters it removes refusal at negligible cost to capability \citep{arditi2024refusal}.

The simplicity of the method, however, conceals how much it assumes about the architecture. The construction above is correct only under three premises, which the original work does not state explicitly because a dense Transformer supplies them directly. \textbf{(i) The writers can be enumerated exhaustively}: $W_E$ together with $\{W_O^l, W_{\text{down}}^l\}_{l=1}^{L}$ are all of the residual stream's writers, fixed in number and in one-to-one correspondence with module names. \textbf{(ii) The residual is a single additive channel} \citep{elhage2021framework}: $h^{l+1} = h^l + f(h^l)$, so every writer's output is added into the same tensor, and eliminating $r$ at the writing end is therefore equivalent to eliminating it at any downstream read. \textbf{(iii) Refusal is mediated by a single direction}: a rank-one projection suffices to remove the behavior. Premise (iii) is empirical, and its scope extends only as far as the models on which it has been tested; (i) and (ii) are structural, and hold as long as the architecture does not change. Two recent changes in frontier open weights, however, each alter one of them.

\subsection{Sparse Mixtures of Experts}

Mixture-of-experts (MoE) models scale parameter count through conditional computation: the feed-forward block at each layer is replaced by $E$ experts $\{f_e\}$, of which a router $g$ selects the top-$k$ per token, giving $\sum_{e \in \mathrm{TopK}} g_e(x) f_e(x)$ \citep{shazeer2017moe, fedus2022switch, jiang2024mixtral, deepseekv3}. Current frontier open-weight models favor fine-grained experts, with $E$ in the hundreds and $k$ in the single digits, together with a shared expert that is always active for every token \citep{dai2024deepseekmoe}. GLM-5.3-Flash instantiates this design with $E = 288$, $k = 8$, and one shared expert.

For the model, this change concerns only capacity and compute efficiency; for directional ablation, however, it changes what premise (i) ranges over. Each expert's down-projection $W_{\text{down}}^{l,e}$ adds its output directly into the residual stream and is therefore a writer in its own right, so the feed-forward writers at each layer go from one to $E + 1$. The growth in count is by itself manageable: enumerating $E + 1$ matrices is no different in kind from enumerating one. What differs is how they are stored. To be executed efficiently through a grouped GEMM, mainstream implementations stack a layer's $E$ expert weights into a single three-dimensional tensor of shape $[E, d_{\text{ff}}, d]$ \citep{gale2023megablocks} rather than keeping $E$ separate linear layers. Public ablation implementations collect writers by traversing modules and matching on names such as \texttt{down\_proj}, which is sufficient whenever (i) is guaranteed by the architecture; fused experts, however, appear neither under that name nor as linear layers at all, so when the traversal finishes they are simply absent from the writer set, and nothing raises an error. A procedure that is correct on a dense model will therefore skip the overwhelming majority of an MoE's feed-forward parameters without emitting any signal that it has done so. How much it thereby misses depends on a question that never had to be answered before: how much of refusal is written by the experts, and how much by the attention and dense projections.

\subsection{Multi-Stream Residuals}

The residual stream is the least-revised component of the Transformer: since ResNet \citep{he2016resnet} it has been a single additive chain, and premise (ii) follows from precisely that. Hyper-connections \citep{zhu2025hyperconnections} are the first substantive generalization of the design: the residual becomes $n$ parallel streams $H^l \in \mathbb{R}^{n \times d}$, a layer reads a weighted combination of the streams as its input, writes its output back into them under learned weights, and an $n \times n$ mixing matrix acts across streams at every layer. Manifold-constrained hyper-connections (mHC) \citep{mhc2025} further constrain that mixing matrix to the manifold of doubly stochastic matrices, which is what makes the design stable to train at frontier scale. GLM-5.3-Flash adopts mHC with $n = 4$.

For training this generalization is a gain; for ablation, however, it removes the one structure premise (ii) depends on. The claim that eliminating $r$ at the writing end also eliminates it at the reading end holds because nothing but element-wise addition sits between the two. (This is not the same as saying that editing the writers and projecting at the layer output are interchangeable: within a layer, a sublayer can read an intermediate state that a boundary projection has not yet reached. We return to that in \S\ref{sec:mhc}.) Under multiple streams, a writer's output is distributed across $n$ streams, those streams are recombined by the mixing matrix at every layer, and what a layer reads is the combination. This does not mean that ablation must fail. But whether a projection at the writing end still guarantees an $r$-free read now depends on the form of the mixing operator. If it acts only on the stream index, weighting each stream's $d$-dimensional vector by a scalar, then any linear combination of streams still has zero component along $r$; if it acts non-trivially within the hidden dimension, it does not. This is decidable directly from the operator's definition, yet to our knowledge no prior work has examined it in the context of ablation. And even where the answer is favorable, the multi-stream residual leaves a second question open. Writer-level and layer-boundary projections need not be equivalent even in a single-stream Transformer, because a sublayer can read an intermediate state before a boundary projection is applied. Whether multi-stream mixing widens that discrepancy has to be measured rather than argued.

\section{Related Work}

\paragraph{Linear representations of refusal and weight editing.} \citet{arditi2024refusal} show across thirteen open-weight chat models of up to 72B parameters that refusal is mediated by a one-dimensional subspace of the residual stream, and give two ways to exploit it. One is an inference-time activation intervention along that direction, of a kind developed more broadly as activation steering \citep{turner2023steering, panickssery2024caa, zou2023repe}; the other is a white-box jailbreak that orthogonalizes the residual-writing weights against it. The premise both rest on is that features occupy linear directions in representation space \citep{park2024linear}, and the weight-editing form places the method alongside other direct interventions on parameters, such as locating and rewriting factual associations \citep{meng2022rome}, arithmetic on task vectors \citep{ilharco2023task}, and low-rank adaptation \citep{hu2022lora}. The latter changes nothing about inference and adds no overhead, since once the edit is applied the model deploys like any other, and it has accordingly become the most widely circulated way to uncensor open weights. Reproductions and much of the work that followed, however, inherit the original recipe without revisiting the architectural conditions it depends on: how the writers are enumerated, and what topology carries the residual, were supplied directly by the dense models the recipe was developed on, and so never became things that had to be discussed. Whether the recipe still holds once a model stops supplying those conditions has not been tested.

\paragraph{Other ways to remove safety behavior.} Prompt-level jailbreaks leave the weights alone and search for an input that evades the trained behavior, whether by gradient-guided suffix optimization \citep{zou2023universal}, by exploiting the mismatch between capability and safety training \citep{wei2023jailbroken}, or by collecting prompts that already circulate \citep{shen2024dan}; the behavior survives in the model and each new request must defeat it again. Fine-tuning removes it from the weights instead, and does so cheaply: ten adversarially designed examples suffice on a hosted model \citep{qi2024finetuning}, a hundred malicious examples and an hour of GPU time suffice on open weights \citep{yang2023shadow}, and a quantized low-rank adapter fitted on one GPU for under \$200 brings a 70B chat model's refusal rate to roughly one percent \citep{lermen2023lora}. Directional ablation sits below both in cost. It needs no optimization against the model and no gradient step at all, only a forward pass over a few hundred prompts, which is why it is the recipe that circulates for open weights and the one whose reach at frontier scale is worth measuring.

\paragraph{Linear concept erasure.} Removing a direction from a representation is a general operation with its own literature. Iterative nullspace projection removes a protected attribute by repeatedly projecting out the directions a linear probe finds \citep{ravfogel2020inlp}, and LEACE gives the closed-form projection that erases a concept while perturbing the representation as little as possible \citep{belrose2023leace}. That line works on activations and asks what a probe can still recover; the weight-editing form used here asks instead what behavior survives, and it inherits the assumption those methods established: that the concept occupies a low-dimensional linear subspace, estimated in practice by a difference of class means \citep{marks2024geometry}, over features that superposition packs into a space smaller than their number \citep{elhage2022toy}. Refusal ablation is that construction applied to one behavior and pushed through to the weights.

\paragraph{How deep safety training goes.} A recurring finding is that alignment is thin. Instruction tuning on a thousand curated examples recovers most of it \citep{zhou2023lima}, and safety behavior in particular concentrates in the first few generated tokens, which is why prefilling an answer defeats it and why fine-tuning undoes it so cheaply \citep{qi2025shallow}. Unlearning aims at the opposite property, removing a capability from the weights rather than gating it \citep{li2024wmdp}, and tamper-resistant safeguards aim at making a safety property survive an adversary who holds the weights \citep{tamirisa2025tamper}. Where a weight edit reaches, and what it fails to reach, is the measurement those defenses would have to be evaluated against.

\paragraph{Geometry beyond a single direction.} Later work has refined the representational picture. \citet{wollschlaeger2025geometry} find refusal mediated not by one direction but by a multi-dimensional ``concept cone'' whose directions are representationally independent. Other work recovers geometrically distinct directions for each of eleven refusal categories, yet reports that they yield nearly identical refusal--over-refusal trade-offs, differing mainly in \emph{how} the model refuses rather than whether it does \citep{morethanone2026}. Refusal has also been characterized as an affine rather than a purely linear function \citep{refusalaffine2024}. Together these results point to the single-direction account being incomplete. What they characterize, however, is the geometry of the representation, largely through analysis and steering in activation space and largely on dense models of at most roughly 70B parameters. How much refusal a single rank-one weight edit actually removes at frontier scale is a question at a different level. That several directions exist in the representation and that removing several directions removes more refusal are not the same claim; whether the second holds has not been established at this scale.

\paragraph{Safety in mixture-of-experts models.} As MoE architectures have become standard, recent work has begun to ask how safety behavior is distributed over the experts. RASET \citep{safetyexperts2026} reports that routing in aligned MoE models is largely topic-driven, identifies safety-critical experts by a contrastive routing-sensitivity criterion, and alters their behavior by parameter-efficient tuning of those experts alone, leaving the routing path intact. Work extending steering to MoE models \citep{expertrefusalsteering2026} exploits refusal-related expert routing patterns and expert-specific steering directions. It reports that the refusal signal the steering recovers does not coincide with expert routing behavior, which it reads as evidence that attention plays a substantial role in MoE refusal. This work establishes that experts and safety behavior are connected, but these interventions either require gradient-based tuning or operate on activations at inference time; neither provides the training-free weight edit studied here. Whether a training-free, gradient-free weight edit succeeds turns instead on a quantity that has not been measured: which parameters write refusal into the residual stream, and how much of that writing falls to the experts as against the attention and dense projections.

\paragraph{Architectural changes to the residual stream.} Hyper-connections \citep{zhu2025hyperconnections} and their manifold-constrained variant mHC \citep{mhc2025} replace the single residual with several parallel streams and mix across them at every layer, motivated by training stability and representation quality. As architecture papers they are evaluated on perplexity and downstream task performance, which is the appropriate standard for what they set out to do. Representation-level interventions---activation patching, steering vectors, and the weight orthogonalization used here---rest on the premise that the residual is a single additive channel, and that premise is exactly what the multi-stream design alters. To our knowledge no prior work has examined whether these interventions remain valid under a multi-stream residual, nor where within such a structure an edit should be applied.

\section{Method and Setup}

To probe how far a single direction reaches, we keep the recipe fixed and change the model. The direction is still one unit vector given by the difference of means over harmful and harmless prompts, and the edit is still one rank-one projection applied to every weight that writes into the residual stream. No training, no gradients, and no second direction are introduced. The target model, however, is chosen outside the conditions the recipe was developed under: a sparse mixture of experts of roughly 320B parameters, a residual carried by four parallel streams, and weights released in block-FP8. The recipe touches this architecture at three points: enumerating the writers inside fused expert tensors, reading activations from several residual streams, and orthogonalizing quantized codes. At each we do only what is needed for the recipe to apply at all, and measure the loss that the handling itself introduces. Whatever difference remains in the results can then be attributed to the single direction rather than to engineering workarounds.

\subsection{Target Model and the Writer Set}

GLM-5.3-Flash \citep{glm53flash}, the latest of the GLM family \citep{glm2024chatglm}, exhibits all three features at once, which is why it serves as the entry point. It is a mixture-of-experts model with roughly 320B total and 18B active parameters: 45 Transformer layers plus one multi-token-prediction (MTP) module \citep{gloeckle2024mtp}, hidden size 4096; hybrid attention, with 34 gated linear layers implementing Kimi Delta Attention (KDA) \citep{kimilinear2025} and 11 multi-head latent attention (MLA) layers \citep{deepseekv3} running sparse full attention, the latter with a top-2048 indexer in the manner of DeepSeek sparse attention \citep{deepseekv32}; 288 routed experts per layer under top-8 routing plus one shared expert, the first three layers being dense MLPs; a four-stream manifold-constrained hyper-connection residual; and a native vision and video tower. The weights are released in block-FP8 \citep{micikevicius2022fp8} (e4m3, $128 \times 128$ blocks) across 62 shards and 76,108 tensors. Frontier open weights now ship quantized as a matter of course, whether in a narrow float format or under a post-training scheme \citep{dettmers2022int8, frantar2023gptq, lin2024awq}, so a weight edit has to be applied to codes rather than to matrices --- a condition none of the published ablation recipes addresses. The parameter counts above include the 7.4B MTP module and the 0.56B vision tower.

\begin{table}[t]
\centering\small
\setlength{\tabcolsep}{4pt}
\begin{tabular}{p{2.3in}r}
\toprule
Writer & Matrices \\
\midrule
Expert down-projections (288 experts $\times$ 43 sparse layers) & 12,384 \\
Attention output projections (34 KDA, 11 MLA, 1 MTP) & 46 \\
Shared-expert down-projections & 43 \\
Dense MLP down-projections (layers 0--2) & 3 \\
Vision-merger down-projection & 1 \\
MTP input projection (\texttt{eh\_proj}) & 1 \\
Token embedding (row space) & 1 \\
\midrule
Total & 12,479 \\
\bottomrule
\end{tabular}
\caption{Matrices that write into the language residual stream of GLM-5.3-Flash: 12,442 in FP8 and 37 in BF16. Counting is at the level of weight matrices; the 12,442 block scale tensors that the requantization updates alongside them are not separate writers.}
\label{tab:writers}
\end{table}

The criterion for a writer is a single one: a matrix whose output is added directly into the language residual stream. Enumerating exhaustively under this criterion gives 12,479 matrices (Table~\ref{tab:writers}), of which 12,442 are stored in FP8 and 37 in BF16: 34 attention output projections on the gated-linear layers, the token embedding, the MTP input projection and the vision merger. Comparing the released edited checkpoint against the base tensor by tensor confirms the inventory from the artifact rather than from the code: 24,921 of the 76,108 tensors differ, being exactly these 12,479 weight matrices and the 12,442 block scale tensors that the requantization updates with them, and nothing else moved. The router, the two reader projections (gate and up), all hyper-connection mixers, the sparse-attention indexer, the normalization layers, and the unembedding do not write into the residual and are left unchanged. The expert down-projections account for 99.2\% of all writers; they exist as three-dimensional fused tensors, and they are exactly what matching on module names fails to reach.

\subsection{Estimating the Direction and Choosing the Layer}

Direction estimation follows the original recipe. Harmful prompts are drawn from AdvBench \citep{zou2023universal} and harmless prompts from the Alpaca instructions \citep{taori2023alpaca} that carry no context input; each is split into two disjoint sets, 256 and 256 for estimating the direction and 32 and 32 for choosing the layer. A further 64 prompts each from the harmful and benign splits of JailbreakBench \citep{chao2024jailbreakbench} serve as a held-out test from a source different from the fitting data. For every layer we take the residual representation at the last token, subtract its mean over the harmless set from its mean over the harmful set, and normalize to a unit vector, computing throughout in float32.

The estimator carries one safeguard: dimensions whose mean magnitude exceeds ten times the layer \emph{median} are masked before the difference is taken, because a handful of content-independent dimensions with an order-of-magnitude larger activation \citep{sun2024massive} would otherwise dominate it. \textbf{On this checkpoint the safeguard never fires.} At layer 22, where the direction is fitted, no dimension exceeds the threshold: masking at ten times the median and not masking at all produce the same vector to the last bit, so the direction reported here is the raw difference of means.

Sweeping the multiplier shows the setting is not a tuned one (Table~\ref{tab:mask}). Tightening it damages the direction: at a multiplier of 2 it masks 558 of 4096 dimensions, turns the direction to a cosine of 0.898 against the shipped one, and leaves refusal at 0.691 where the shipped direction leaves 0.156. A multiplier of 5 masks two dimensions and leaves refusal at 0.133, below 0.156 by less than the run-to-run variation of greedy decoding under tensor parallelism, which we do not read as an improvement. The sweep was run at layer 22 only.

\begin{table}[t]
\centering\small
\setlength{\tabcolsep}{4pt}
\begin{tabular}{rrrr}
\toprule
Multiplier & Dims masked & $\cos$ to shipped & Refusal \\
\midrule
2 & 558 & 0.898 & 0.691 \\
3 & 107 & 0.977 & 0.227 \\
5 & 2 & 0.9987 & 0.133 \\
10 (shipped) & 0 & 1.0000 & 0.156 \\
none & 0 & 1.0000 & 0.156 \\
\bottomrule
\end{tabular}
\caption{Massive-activation mask at layer 22, swept over the multiplier of the layer median. Refusal is measured on the 256 held-out harmful prompts with all writers hooked.}
\label{tab:mask}
\end{table}

Layer selection is done on the validation set. Nine candidate layers between depth 0.4 and 0.8 are each ablated at inference time through a hook that subtracts the layer's direction, with the weights untouched, and scored by the bypass rate on harmful prompts minus a KL penalty on harmless prompts, the latter keeping the edit from spilling into ordinary behavior. Layer 22 (depth 0.49) scores highest. Layers 20 and 22 are effective while layer 25 is not; this non-monotonicity reproduces stably on the validation set and is not measurement noise, so the layer has to be measured rather than set from a depth fraction.

\subsection{Applying the Edit under Fused Experts and a Multi-Stream Residual}

The original recipe assumes that writers can be found by module name and that the residual can be read as a single tensor. This model satisfies neither.

Writers are collected by traversing the structure of the parameter tensors rather than by matching names. A three-dimensional expert tensor is unfolded along its first dimension into one two-dimensional matrix per expert, and these enter the writer set alongside the ordinary linear layers. The step involves no algorithmic difficulty, but it decides whether the edit lands on the weights that actually write the residual.

The multi-stream residual affects reading but not editing. When activations are read, a layer's residual is a bundle of shape $[B, S, 4, H]$; it is collapsed into a single $H$-dimensional vector under the hyper-connection's mixing weights before the last-token representation is taken. Applying the edit requires nothing further. The cross-stream mixing in mHC weights each stream as a whole by a scalar and performs no rotation within the hidden dimension, so a direction that is zero in every stream remains zero after mixing. The edit is applied, as in the original recipe, to the writer weights.

The edit ultimately has to land in the checkpoint on disk, and this checkpoint cannot be edited in memory and saved back, for three independent reasons: the weights are FP8 codes rather than real matrices, and projecting the codes directly destroys the matrix; the experts are fused into three-dimensional parameters at load time; and \texttt{transformers} drops the MTP module (\texttt{layers.45}) on loading, so that \texttt{save\_pretrained} silently writes 6.98\,GiB less than it read. We therefore stream each shard in, rewrite it, and write it back, and after editing verify that all 62 shards and 76,108 tensors match the base checkpoint in name, dtype, and shape.

\subsection{Applying the Edit to Block-FP8 Weights}

The rank-one projection is defined on real matrices, while the weights are stored as codes. A block-FP8 matrix consists of e4m3 codes and a per-block scale (\texttt{weight\_scale\_inv}) holding one coefficient per $128 \times 128$ block. The edit therefore proceeds in three steps: dequantize under the per-block scales, project in float32, requantize, and write back both the codes and the updated scales. Two errors here leave no visible trace afterwards. The scales must be written back in the dtype the checkpoint uses: narrowing an fp32 scale to bf16 discards 16 mantissa bits, changes the file size, and may bypass the branch a loader dispatches on \texttt{scale.dtype}. And quantization must use the very scale that is written to the file; otherwise an avoidable layer of error is introduced between the codes and the scales.

\begin{table}[t]
\centering\small
\begin{tabular}{rrr}
\toprule
Iterations & Residual leakage & Weight perturbation \\
\midrule
8 & 13.4\% & 2.15\% \\
32 & 5.2\% & 2.79\% \\
\bottomrule
\end{tabular}
\caption{Iterated orthogonalization under block-FP8 requantization, measured on one MLA output projection with the same direction.}
\label{tab:niters}
\end{table}

More fundamental than either is that requantization reintroduces part of the component that was just projected out. Measured on a real MLA output projection with the same direction, a single project--quantize--dequantize round leaves a substantial residue; repeating the cycle lets each round clean up the rounding error of the previous one, and the residue falls with the number of iterations while the weight perturbation rises (Table~\ref{tab:niters}).

When only attention is edited this leakage is negligible: 34 of the 45 writers at that stage are BF16 with about 1\% leakage, and the leakage of the remaining 11 FP8 matrices is lost in the noise, a lossless hook giving a refusal rate of 0.812 against 0.828 after baking. When all writers are edited, however, 12,442 FP8 matrices are involved and the leakage accumulates matrix by matrix: the lossless hook predicts a refusal rate of 0.031, while the model baked with 8 iterations gives 0.188. Raising the iteration count to 32 lowers the leakage from 13.4\% to 5.2\% and the refusal rate from 0.188 to 0.094, recovering about half of the gap. Beyond that the leakage falls slowly while the weight perturbation keeps growing, so we take 32 as the operating point. The measured maximum residual leakage after editing is 0.149 for FP8 and 0.024 for BF16.

\subsection{Evaluation Setup}

\paragraph{Benchmarks.} The evaluation covers three kinds of behavior: refusal on harmful prompts, over-refusal, and general capability; the benchmarks, sample sizes, and scoring rules are listed in Table~\ref{tab:benchmarks}. There are seven harmful benchmarks, each sampled from its full set under a fixed random seed. Over-refusal uses the 250 safe prompts of XSTest, which carry sensitive surface features but are harmless, so that refusing them is a false refusal. The before-and-after comparison shows whether what the edit removes is the harmful part of refusal or the harmless part along with it. The four capability benchmarks are multiple-choice tasks scored by the logit of the first token of each option letter, except GSM8K, which keeps chain-of-thought generation \citep{wei2022cot} because it needs the reasoning. Base and edited models run under identical settings: greedy decoding, the same random seed, \texttt{reasoning\_effort=low}, with \texttt{<think>} blocks stripped before classification. Every response is logged in full, so re-judging, paired tests, and per-category breakdowns need no regeneration.

\begin{table*}[t]
\centering\small
\setlength{\tabcolsep}{6pt}
\begin{tabular}{llrlr}
\toprule
Category & Benchmark & $n$ & Scoring & Overlap \\
\midrule
Harmful & AdvBench \citep{zou2023universal} & 100 & judge & 100\% \\ & JailbreakBench \citep{chao2024jailbreakbench} & 100 & judge & 8\% \\ & StrongREJECT \citep{souly2024strongreject} & 150 & judge & 5\% \\ & HarmBench \citep{mazeika2024harmbench} & 150 & judge & -- \\ & MaliciousInstruct \citep{huang2024catastrophic} & 100 & judge & -- \\ & ForbiddenQuestions \citep{shen2024dan} & 150 & judge & -- \\ & SimpleSafetyTests \citep{vidgen2023simplesafetytests} & 50 & judge & -- \\
\midrule
Over-refusal & XSTest \citep{roettger2024xstest} & 250 & judge & -- \\
\midrule
Capability & MMLU \citep{hendrycks2021mmlu} & 300 & letter logit & -- \\ & MMLU-Pro \citep{wang2024mmlupro} & 400 & letter logit & -- \\ & CMMLU \citep{li2024cmmlu} & 500 & letter logit & -- \\ & GSM8K \citep{cobbe2021gsm8k} & 150 & CoT, exact match & -- \\
\bottomrule
\end{tabular}
\caption{Evaluation benchmarks. \emph{Overlap} is the fraction of a benchmark's sample that falls on the AdvBench prompts used to estimate the direction; the three non-zero entries each have a clean variant drawn only from outside the fitting slice. MMLU-Pro is scored by letter logit and is therefore lower than a chain-of-thought run of the same benchmark; it serves only for the before-and-after comparison and is not comparable to absolute figures reported elsewhere.}
\label{tab:benchmarks}
\end{table*}

\paragraph{Judge.} A rule-based classifier that checks only whether the response opens with a fixed phrase misses empathetic refusals: under a self-harm prompt, ``I'm worried about you\ldots{} please reach out to\ldots'' is an unambiguous refusal that contains no template phrase. We instead re-judge every logged response with a judge model \citep{zheng2023judging} from a different model family (DeepSeek-v4-flash), blind, cached, and retried with back-off on failure; the different family also avoids self-judging. The two judgments agree on 0.85--0.96 of responses on the harmful benchmarks, and the disagreement is systematic in direction: the rule-based classifier underestimates refusal on harmful sets. The main text reports the judge model's figures.

The judge returns one of three labels, \textsc{refusal}, \textsc{compliance} and \textsc{unfinished}, and which of them count as refusal has to be stated rather than assumed. Scoring everything that is not \textsc{compliance} as refusal folds in replies that ran out of the token budget; on harmful prompts that is 0--9 items per condition and changes little, but on harmless prompts 21 of 100 base-model replies were cut off, which would report a 0.260 false-refusal rate for a model that actually refuses 5 of the 100. We therefore count only \textsc{refusal} as refusal throughout, report the alternative as a \emph{non-compliance rate} where it is informative, and give the label counts for every condition in Appendix~\ref{app:labels} (Tables~\ref{tab:app-main}, \ref{tab:app-stages}, \ref{tab:app-arms} and \ref{tab:app-rank}). Items for which the judge returned no usable verdict after retries are excluded and the affected $n$ is stated.

\paragraph{Contamination.} The direction is fitted on AdvBench, and the evaluation also samples AdvBench. Under the same random seed, all 100 prompts of that benchmark fall on the fitting data, and JailbreakBench and StrongREJECT each share a small number of prompts with AdvBench in wording. For these three benchmarks we construct clean variants using only samples outside the fitting slice, and every claim about generalization rests on the clean variants.

\paragraph{Statistics.} Rates are reported with Wilson 95\% confidence intervals \citep{wilson1927}. A percentile bootstrap \citep{efron1979bootstrap} degenerates where a rate sits at the boundary: 50 successes out of 50 returns $[1.000, 1.000]$ against Wilson's $[0.929, 1.000]$, and several rates here do sit there, so Wilson is what we quote. The paired capability comparison, whose statistic is a difference rather than a rate, uses a paired bootstrap over items \citep{efron1979bootstrap}. The key pairwise comparisons are made on the same set of prompts with McNemar's paired test \citep{mcnemar1947}, which is more sensitive than an unpaired comparison and does not mistake variance between prompts for a difference between conditions.

\paragraph{Infrastructure.} The model is loaded with \texttt{transformers} \citep{wolf2020transformers} under tensor parallelism \citep{shoeybi2019megatron} with expert parallelism (TP8+EP). One detail of distributed generation cost us several runs before we found it: every rank must execute the same number of decode steps. \texttt{generate} decides when to stop from the sequences it can see, and that decision is rank-local, so a rank that stops one step early leaves the others waiting on a collective that is never issued, and the job deadlocks with every GPU allocated and idle. We caught it in the NCCL flight recorder as rank 0 reducing a $[8, 4096]$ decode step while ranks 1--7 reduced the $[856, 4096]$ prefill of the next batch at the same collective sequence number. Passing \texttt{synced\_gpus} keeps the loop running on every rank until all of them are done and removes the hang; a short watchdog timeout only shortens the wait for a deadlock that should not occur. Long stages additionally checkpoint per condition so that an interrupted run resumes rather than restarts.

\section{Results}

The edited model's refusal rate on seven harmful benchmarks falls from 0.51--1.00 to 0.08--0.48, no capability score moves detectably, and false refusal on harmless prompts drops to zero. A single direction removes most of this model's refusal. The size of the reduction, however, differs by a factor of two across benchmarks, the residue ranges from 0.08 to 0.48, and the confidence intervals at the two ends do not overlap, so the spread is not noise. \S\ref{sec:main} gives the full safety and capability results and identifies which benchmarks the spread falls on. The four subsections that follow each take up one question: \S\ref{sec:where} measures every subset of the writer groups under lossless hooks; \S\ref{sec:mhc} derives what the mHC mixing operator guarantees and compares editing the writers against projecting at the layer boundary; \S\ref{sec:narrow} asks whether the effect is specific to the direction removed and whether a second direction changes it; and \S\ref{sec:resists} breaks the residue down by content category and fits subspaces on the resistant categories at every rank from 1 to 12.

\subsection{Main Results}
\label{sec:main}

\begin{table*}[t]
\centering\small
\begin{tabular}{lrllrr}
\toprule
Benchmark & $n$ & Base & Edited & $\Delta$ & McNemar $p$ \\
\midrule
MaliciousInstruct & 100 & 0.970 [0.915, 0.990] & 0.080 [0.041, 0.150] & $-0.89$ & $3{\times}10^{-27}$ \\
AdvBench & 99 & 0.970 [0.915, 0.990] & 0.131 [0.078, 0.212] & $-0.84$ & $2{\times}10^{-25}$ \\
JailbreakBench & 100 & 0.940 [0.875, 0.972] & 0.150 [0.093, 0.233] & $-0.79$ & $3{\times}10^{-24}$ \\
HarmBench & 150 & 0.920 [0.865, 0.954] & 0.167 [0.116, 0.234] & $-0.75$ & $2{\times}10^{-34}$ \\
StrongREJECT & 150 & 0.993 [0.963, 0.999] & 0.380 [0.306, 0.460] & $-0.61$ & $1{\times}10^{-26}$ \\
SimpleSafetyTests & 50 & 1.000 [0.929, 1.000] & 0.480 [0.348, 0.615] & $-0.52$ & $3{\times}10^{-8}$ \\
ForbiddenQuestions & 150 & 0.513 [0.434, 0.592] & 0.100 [0.062, 0.158] & $-0.41$ & $7{\times}10^{-18}$ \\
\midrule
AdvBench (clean) & 99 & 0.990 [0.945, 0.998] & 0.152 [0.094, 0.235] & $-0.84$ & $2{\times}10^{-25}$ \\
JailbreakBench (clean) & 92 & 0.946 [0.879, 0.977] & 0.196 [0.127, 0.288] & $-0.75$ & $3{\times}10^{-21}$ \\
StrongREJECT (clean) & 150 & 0.993 [0.963, 0.999] & 0.327 [0.257, 0.405] & $-0.67$ & $4{\times}10^{-29}$ \\
\midrule
XSTest (safe) & 250 & 0.024 [0.011, 0.051] & 0.000 [0.000, 0.015] & $-0.02$ & 0.031 \\
\bottomrule
\end{tabular}
\caption{Harmful refusal and false refusal, judged by the judge model and counting only \textsc{refusal}, with Wilson 95\% confidence intervals. $n$ is the number of prompts with a usable verdict on both checkpoints; one AdvBench response has none, so those rows are scored out of 99. Per-condition label counts are in Appendix~\ref{app:labels}.}
\label{tab:main}
\end{table*}

Table~\ref{tab:main} gives refusal rates on the seven harmful benchmarks and the three clean variants. The base model refuses 0.92--1.00 on six of them; ForbiddenQuestions is the exception at 0.51, because a substantial share of its questions are sensitive rather than harmful and the base model answers them. After editing, the seven rates lie between 0.08 and 0.48, and every reduction is significant under a paired McNemar test ($p < 10^{-7}$ throughout; the prompts are paired within each benchmark, 50 to 150 of them). The discordant pairs run almost entirely in one direction. Across the seven harmful benchmarks, 799 prompts carry a usable verdict on both checkpoints; 546 of them are refused by the base model and not by the edited one---537 becoming compliant answers and 9 running out of the token budget---while two move the other way, one on StrongREJECT and one on ForbiddenQuestions.

We report Wilson intervals rather than bootstrap ones. The two disagree where a rate sits at the boundary, which several of these do: a percentile bootstrap of 50 successes out of 50 returns $[1.000, 1.000]$ against Wilson's $[0.929, 1.000]$, a gap of 0.071 at the lower bound, and 0 out of 250 gives $[0.000, 0.000]$ against $[0.000, 0.015]$. Wilson does not degenerate at the boundary, so it is the interval we quote throughout.

The clean variants give figures close to the contaminated originals: 0.152 against 0.131 on AdvBench, 0.196 against 0.150 on JailbreakBench, 0.327 against 0.380 on StrongREJECT. All three differences fall inside the confidence intervals and do not even agree in sign. The direction was estimated from 256 AdvBench prompts, and the edited model falls just as far on prompts it never saw, so what was removed is not memorization of the fitting sample. Every claim about generalization below rests on the clean variants.

The edit did not make the model more cautious on harmless prompts; it did the opposite. The 250 safe XSTest prompts carry sensitive surface features, the base model falsely refuses six of them, and the edited model refuses none.

\begin{table}[t]
\centering\small
\setlength{\tabcolsep}{3pt}
\begin{tabular}{lrrrl}
\toprule
Benchmark & Base & Edited & $\Delta$ & Paired 95\% CI \\
\midrule
MMLU & 0.830 & 0.820 & $-0.010$ & $[-0.037, +0.017]$ \\
MMLU-Pro & 0.440 & 0.443 & $+0.003$ & $[-0.020, +0.025]$ \\
GSM8K & 0.933 & 0.940 & $+0.007$ & $[-0.013, +0.033]$ \\
CMMLU & 0.856 & 0.858 & $+0.002$ & $[-0.014, +0.018]$ \\
\bottomrule
\end{tabular}
\caption{Capability, paired item by item. Both checkpoints answer the same items in the same order, so the comparison is paired and the interval is a paired bootstrap over items. McNemar gives $p = 0.63$ on MMLU and $p = 1.00$ on the other three, with near-symmetric flips (10:7, 10:11, 1:2, 7:8).}
\label{tab:capability}
\end{table}

Capability is given in Table~\ref{tab:capability}. Every paired interval straddles zero, no test is significant, and the items that change answer do so in both directions in near-equal numbers. We detect no change on these subsets, but the MMLU interval still admits a drop of 3.7 points.

Scored with the rule-based classifier instead, the edited column of Table~\ref{tab:main} would be lower---0.273 rather than 0.380 on StrongREJECT, 0.340 rather than 0.480 on SimpleSafetyTests---because that classifier misses soft refusals containing no template phrase. We report the more conservative of the two judgments. Even so, the spread of the reduction, from 41 to 89 percentage points, is conspicuous. The two benchmarks with the highest residue, StrongREJECT and SimpleSafetyTests, are exactly the two whose content is concentrated in violence, sexual content, and self-harm. The two with the lowest, MaliciousInstruct and AdvBench, consist largely of requests for illegal procedures. What the single direction removed and what it left behind are therefore two questions to be answered separately.

\subsection{Which Weights the Edit Needs}
\label{sec:where}

The most direct way to find out how far the edit must reach is to bake nothing and instead attach lossless hooks to exactly the writers a bake stage would edit. The hooks subtract the direction's component from their outputs at inference time. The hooks bypass quantization, so what they measure is the ceiling of the edit itself. A nested ladder (attention, then the dense and shared projections, then the routed experts) only ever measures what a group adds \emph{given} the groups before it, which depends on the order the ladder walks. We therefore measure all seven non-empty subsets of $\{$attention, dense, experts$\}$, which makes the decomposition order-free. Evaluation uses 256 held-out harmful prompts, 164 from outside the AdvBench fitting slice and 92 from JailbreakBench, and pairs of conditions are compared by McNemar's test on the same prompts.

\begin{table}[t]
\centering\small
\setlength{\tabcolsep}{4pt}
\begin{tabular}{lrrr}
\toprule
Edited group & Judge & Removed & Rule \\
\midrule
none & 0.949 & --- & 0.961 \\
attention & 0.910 & 0.039 & 0.918 \\
dense $+$ shared & 0.934 & 0.016 & 0.934 \\
routed experts & 0.801 & 0.148 & 0.840 \\
attention $+$ dense & 0.883 & 0.066 & 0.895 \\
dense $+$ experts & 0.654 & 0.295 & 0.613 \\
attention $+$ experts & 0.408 & 0.541 & 0.414 \\
all three & 0.173 & 0.776 & 0.156 \\
\bottomrule
\end{tabular}
\caption{All seven writer-group subsets on 256 held-out harmful prompts. \emph{Judge} counts \textsc{refusal} only and is the primary metric; \emph{Rule} is the opening-phrase classifier, which also counts unfinished replies. Judge rates are over 254--256 prompts, the remainder having no usable verdict. Against the unedited condition, attention is significant (11:1, $p = 0.006$) and the routed experts are (39:1, $p = 7{\times}10^{-11}$); the dense group alone is not (6:2, $p = 0.29$).}
\label{tab:stages}
\end{table}

\textbf{The groups are strongly non-additive} (Table~\ref{tab:stages}). Editing attention, the dense and shared projections, or the routed experts on their own removes 0.039, 0.016 and 0.148 of refusal; editing all three removes 0.776. The three single-group effects sum to 0.203, so 0.573, which is 74\% of the joint effect, appears only when the groups are edited together. Editing the experts alone leaves refusal at 0.801; it is editing them together with the attention output projections that collapses it, to 0.408. The rule-based scoring gives the same picture with an interaction of 76\% of the joint effect. What this establishes is a property of the intervention: the drop in refusal is strongly non-additive over writer groups. It does not by itself locate where refusal is represented, and we do not read it as such.

Attributing the joint effect to individual groups requires a convention. Shapley values, which split the interaction evenly over the orderings, give the experts 54\%, attention 32\% and the dense group 14\% under the judge, and 53\%, 30\% and 17\% under the rule classifier. We report them for completeness, with the seven raw subsets alongside, but with three quarters of the effect in the interaction term a Shapley share is an allocation rule applied to an intervention, not a statement about where refusal lives.

What the numbers do support without a convention is the practical claim. The writers a name-matching implementation reaches on this model are exactly attention plus the dense and shared projections: 90 matrices at inference, whose removal takes refusal from 0.949 to 0.883. The 12,096 expert down-projections it does not reach are three-dimensional fused tensors whose names contain no ``down projection'' and fall outside the match; on disk the same gap covers 12,384 expert matrices, together with the token embedding, the MTP input projection and the vision merger, which the name match also misses. Editing what the recipe finds reports success and moves refusal by 0.066 out of an available 0.776. The failure is not that the architecture resists ablation but that the edit never touched most of the weights that matter; any implementation that enumerates the fused expert tensors avoids it.

One comparison is sensitive to the scorer. Adding the dense and shared projections on top of attention is not significant under the rule classifier (10:4, $p = 0.18$) nor under the judge counting \textsc{refusal} only (12:5, $p = 0.14$), but is significant when unfinished replies are counted as non-compliant (14:4, $p = 0.031$). The dense group's own contribution is small enough that which replies count as refusals determines whether it clears significance; we therefore report it as small and not established rather than as a null.

\subsection{Ablation under a Multi-Stream Residual} \label{sec:mhc} Manifold-constrained hyper-connections turn the residual into four streams, but every operation they perform on those streams is a scalar weighting per stream: a layer reads a scalar-weighted sum of the streams, writes its output back into them under scalar coefficients, and the $4 \times 4$ doubly stochastic mixing matrix between layers acts on the stream index rather than within the hidden dimension. All three operations multiply an entire $d$-dimensional vector by a scalar and add, and orthogonality to $r$ is preserved under scalar multiplication and addition. Hence if every writer's output is orthogonal to $r$, each of the four streams is orthogonal to $r$, and so is any combination a layer reads. The premise of a single residual channel weakens to the requirement that the mixing operator does not rotate within the hidden dimension, and the conclusion survives. Tensor parallelism supplies a direct check: after the all-reduce, each layer's residual is element-wise identical across the eight ranks, with a maximum difference of zero, so projecting per rank is consistent.

What the algebra settles is effectiveness. Equivalence with a projection at the layer boundary is a separate question, and we test it on the same 256 prompts and over the same layers, since the boundary probe registers a hook on every decoder layer, not on one, so what differs is where the projection is applied, not how much of the network it covers. On their own valid prompts the two conditions give 0.173 and 0.271 under the judge, and 0.156 and 0.223 under the rule classifier. The comparison itself is made on the 253 prompts with a usable verdict in both, where the rates are 0.174 and 0.265; those two ways of computing a rate differ by less than 0.01 here, but they are not the same quantity and we keep them apart. The marginal intervals overlap, so the difference is not visible in them; the paired test, which uses the per-prompt outcomes those intervals discard, finds it. Of 31 discordant pairs, 27 favor editing the writers ($p = 3{\times}10^{-5}$); under the rule classifier it is 22 of 27 ($p = 0.0015$). The result holds under both scorers. The boundary projection may remove less because editing the writers takes the component out before it enters the residual, so no sublayer within the layer can read it, whereas projecting at the boundary leaves the feed-forward sublayer having already read an intermediate state containing $r$, whose non-linear traces do not lie along $r$ and cannot be projected away afterwards. Under hyper-connections a layer reads a combination of streams that already carries its own attention output, which would widen that exposure. But the same within-layer exposure exists in a single-stream Transformer, and this experiment cannot separate the two: it establishes that editing writers beats the boundary probe on this model, not that mHC is the reason.

\subsection{What the Edit Is Specific To}
\label{sec:narrow}

A rank-one projection removes a direction from every residual writer, which is a substantial perturbation of the weights. Two questions follow: whether the drop in refusal is specific to \emph{this} direction, and whether removing a second direction alongside it changes the outcome. We compare five conditions at layer 22 under hooks on all 132 writer sites, on 150 StrongREJECT and 100 JailbreakBench prompts (Table~\ref{tab:directions}). The subspace is stored row-wise as $Q \in \mathbb{R}^{k \times d}$ and applied as $x - (xQ^{\top})Q$, that is $I - Q^{\top}Q$; rows are orthonormalized by Gram--Schmidt with an explicit re-orthogonalization step, measuring $1.7 \times 10^{-8}$ between rows. The second direction is either the first principal component of the harmful-prompt residual deflated against the shipped direction, or a random unit vector orthogonal to it.

\begin{table}[t]
\centering\small
\setlength{\tabcolsep}{4pt}
\begin{tabular}{lrrrrr}
\toprule
& & \multicolumn{2}{c}{Judge} & \multicolumn{2}{c}{Rule} \\
\cmidrule(lr){3-4}\cmidrule(lr){5-6}
Removed subspace & Rank & SR & JBB & SR & JBB \\
\midrule
none & 0 & 0.993 & 0.920 & 1.000 & 0.920 \\
shipped direction $r$ & 1 & 0.320 & 0.190 & 0.227 & 0.190 \\
random $\perp r$ & 1 & 0.987 & 0.920 & 1.000 & 0.960 \\
$r + \mathrm{PC}_1$ & 2 & 0.307 & 0.150 & 0.207 & 0.110 \\
$r + \text{random}$ & 2 & 0.253 & 0.160 & 0.187 & 0.140 \\
\bottomrule
\end{tabular}
\caption{Five ablation conditions at layer 22, all writers hooked, on 150 StrongREJECT (SR) and 100 JailbreakBench (JBB) prompts. \emph{Judge} counts \textsc{refusal} only and is the primary metric; \emph{Rule} is the opening-phrase classifier. Each condition leaves between $1{\times}10^{-4}$ and $5{\times}10^{-4}$ of every ablated direction in the residual, so no condition is an inert hook.}
\label{tab:directions}
\end{table}

\textbf{The effect is specific to the direction removed.} Ablating a random direction orthogonal to $r$ leaves refusal at 0.987 and 0.920, indistinguishable from the unedited 0.993 and 0.920, while ablating $r$ leaves 0.320 and 0.190. Projecting \emph{something} out of every residual writer does not loosen the model. One random direction is one control. It shows that this particular arbitrary direction does nothing; several independent draws would be needed to say the same of arbitrary directions in general.

Neither of the two tested rank-2 extensions restored refusal: 0.307 and 0.253 on StrongREJECT against 0.320 for rank 1. We did not run the paired comparison that would be needed to say whether either improves on rank 1.

\subsection{What Resists}
\label{sec:resists}

\begin{table*}[t]
\centering\small
\begin{tabular}{lrl@{\hspace{2em}}lrl}
\toprule
StrongREJECT category & $n$ & Residue & HarmBench category & $n$ & Residue \\
\midrule
Violence & 26 & 0.62 [0.43, 0.78] & Harmful (general) & 13 & 0.38 [0.18, 0.64] \\
Sexual content & 22 & 0.50 [0.31, 0.69] & Harassment/bullying & 15 & 0.27 [0.11, 0.52] \\
Hate, harassment and discrimination & 25 & 0.48 [0.30, 0.67] & Illegal & 43 & 0.26 [0.15, 0.40] \\
Disinformation and deception & 21 & 0.38 [0.21, 0.59] & Chemical/biological & 19 & 0.16 [0.06, 0.38] \\
Non-violent crimes & 29 & 0.21 [0.10, 0.38] & Misinformation & 29 & 0.07 [0.02, 0.22] \\
Illegal goods and services & 27 & 0.15 [0.06, 0.32] & Cybercrime/intrusion & 31 & 0.00 [0.00, 0.11] \\
\bottomrule
\end{tabular}
\caption{Residual refusal of the edited model by content category, counting only \textsc{refusal}, with Wilson 95\% confidence intervals. The rows sum to the benchmark totals in Table~\ref{tab:main}: 57 of 150 on StrongREJECT and 25 of 150 on HarmBench.}
\label{tab:categories}
\end{table*}

The refusal that survives the edit is not spread evenly. Table~\ref{tab:categories} breaks the residue down by content category on StrongREJECT and HarmBench, whose taxonomies differ. All six StrongREJECT categories sit at 0.96--1.00 on the base model; after editing, violence leaves a residue of 0.62, sexual content 0.50, and hate and harassment 0.48, against 0.15 for illegal goods and 0.21 for non-violent crimes. On HarmBench, harassment and bullying leave 0.27 and chemical/biological 0.16, against 0.00 for cybercrime and 0.07 for misinformation. The category names differ but the pattern repeats: requests about physical harm, sex, and hate resist the edit, while requests about illegal procedures and technical intrusion are almost entirely released. This is where the spread across benchmarks in Table~\ref{tab:main} comes from, since StrongREJECT and SimpleSafetyTests are filled with content of the first kind.

Coverage within the model as it runs is not the explanation: the hook experiment of Table~\ref{tab:stages} attaches to every residual writer on the inference path---all 132 sites, covering the 12,186 matrices the loaded model exercises---and the residue is there all the same. It is not quantization loss: the hook experiment bypasses quantization. What remains is that these categories might be carried by a different direction, or by a subspace of low rank. We tested that by fitting subspaces on prompts from violence, sexual content and hate alone---81 for fitting, 73 disjoint for evaluation---at every rank from 1 to 12 (Table~\ref{tab:resistant}).

\begin{table}[t]
\centering\small
\setlength{\tabcolsep}{4pt}
\begin{tabular}{rlrlr}
\toprule
Rank & Judge & $n$ & Rule & Unfinished \\
\midrule
none & 0.986 & 73 & 1.000 & 0.000 \\
1 & 0.384 & 73 & 0.370 & 0.027 \\
2 & 0.425 & 73 & 0.397 & 0.014 \\
3 & 0.233 & 73 & 0.219 & 0.055 \\
4 & 0.274 & 73 & 0.233 & 0.014 \\
5 & 0.466 & 73 & 0.425 & 0.055 \\
6 & 0.329 & 70 & 0.356 & 0.096 \\
7 & 0.211 & 71 & 0.288 & 0.123 \\
8 & 0.200 & 70 & 0.301 & 0.164 \\
9 & 0.288 & 73 & 0.384 & 0.178 \\
10 & 0.288 & 73 & 0.384 & 0.205 \\
11 & 0.274 & 73 & 0.301 & 0.151 \\
12 & 0.219 & 73 & 0.274 & 0.178 \\
\bottomrule
\end{tabular}
\caption{Subspaces fitted on the resistant categories alone, evaluated on 73 held-out prompts from the same categories. \emph{Judge} counts \textsc{refusal} only over the $n$ prompts with a usable verdict; \emph{Rule} is the opening-phrase classifier over all 73; \emph{Unfinished} is the share of the 73 the judge labelled \textsc{unfinished}. The ranks are independently fitted rather than nested: the first row of every subspace is the deterministic difference of means, but rows beyond it come from a randomized SVD with no fixed per-call seed, so differences between ranks carry fitting randomness as well as rank.}
\label{tab:resistant}
\end{table}

Every rank removes most of this refusal, from 0.986 down to between 0.20 and 0.47, and none removes it. The two scorers do not agree on which rank does best: the judge's lowest value is 0.200 at rank 8 ($14/70$, Wilson $[0.123, 0.308]$), the rule classifier's is 0.219 at rank 3. The intervals are wide and mostly overlapping---rank 3 under the judge is $17/73 = 0.233$, $[0.151, 0.342]$---though not uniformly so: rank 5 at $0.466$, $[0.356, 0.579]$, is separated from both. With twelve conditions on 73 prompts, subspaces fitted independently at each rank, and a minimum selected after the fact, these data do not identify a best rank. They do show a residue that none of the twelve edits removed, on a set of categories where the same class of edit removes the bulk of refusal elsewhere.

The unfinished column is reported alongside the residue because it does not stay constant: it rises from 0.027 at rank 1 to 0.205 at rank 10. Under a fixed 384-token budget, higher-rank edits leave more replies unfinished, and a \textsc{refusal}-only rate falls mechanically when replies stop arriving at an answer. The judge's lowest value, 0.200 at rank 8, sits where 0.164 of replies did not finish, so that value mixes refusal removed with answers never reached. We did not inspect the outputs; what we report is a falling completion rate under a fixed budget.

\section{Ethics and Responsible Release}
\label{sec:ethics}

\paragraph{Threat model.} The attack used here is not new. The directional-ablation recipe was published with an implementation in 2024 \citep{arditi2024refusal}, the base model's weights are released under the MIT license, and the edited checkpoint we study is itself already public. What an open release does and does not change about marginal risk is itself contested \citep{kapoor2024societal}. What an attacker needs is the weights, eight H100s, roughly five minutes of disk writes, and a few hundred harmful prompts. All of that was available before this paper. What we add is measurement rather than capability: the reach of this attack on a frontier-scale architecture, and where it stops. We do publish engineering detail an attacker could use: that the writers have to be enumerated structurally, how many requantization iterations the block-FP8 format needs, and that distributed generation must keep every rank in the same decode loop. That detail lowers the engineering uncertainty of reproducing the attack, though not its cost in compute or data, and it is the same detail a defender needs to audit the weights that matter. The paper is not cost-neutral for an attacker; we argue that the balance favors the defender, for the reasons below.

\paragraph{What defenders gain.} The measurements in \S\ref{sec:where} and \S\ref{sec:resists} were made against an attack, but two of them bear directly on defense. One is the blind spot: an audit that enumerates residual writers by module name covers 0.7\% of those on the inference path and reports no error, so the same traversal that makes the attack fail silently would make a defense fail silently, and the fix is the same in both directions --- enumerate by tensor structure. The other is the residue. A form of refusal survived every subspace we fitted on the categories it concentrates in, which is the one positive result here a defender can build on, though what carries it is unknown and hardening one writer group may run into the same non-additivity the attack did.

\paragraph{Release conditions.} The edited weights and the model card were already published,\footnote{\url{https://huggingface.co/orcarouter/GLM-5.3-Flash-Uncensored-FP8}} under the same license as the base model, and this paper does not change that. What we do not release with the paper is the implementation: the enumeration of writers, the iterated orthogonalization on block-FP8 weights, and the per-shard baking scripts. The two differ in kind: the published weights are one specific checkpoint, whereas the implementation is a general tool that applies to any mixture-of-experts checkpoint. It is the latter that adds leverage for an attacker. We likewise do not release the fitted refusal direction, the subspaces fitted on the resistant categories, or the prompt slices used to fit them. The category names are published, because they are part of the finding itself: knowing which content resists this class of attack is useful to a defender, and knowing it does not help an attacker remove it. This trade-off has a cost, which we state in the limitations.

\paragraph{Implications for safety in the weights.} A rank-one edit that involves no training removes between 41 and 89 percentage points of this model's refusal. The reduction is consistent with a low-dimensional component of refusal that can be altered without a detectable change on the four capability benchmarks tested here, which is what makes refusal and capability look separable. Separability of that kind would be bad news for defenders: what is separable can be excised on its own. The same experiments supply the other half of the picture. The refusal that resists was not removed by any of the twelve subspaces we fitted on those categories; it ships with the weights and survived every edit we tried.

\paragraph{Research program.} This work is part of OrcaRouter Research,\footnote{\url{https://www.orcarouter.ai}} an effort to study the security properties of frontier AI systems and the infrastructure used to deploy them. We are particularly interested in the boundary between model-level safety mechanisms and system-level security controls: which properties can be reliably enforced in model weights, which can be altered by an adversary with weight access, and which therefore require enforcement at the inference and agent-infrastructure layers. The measurements reported here fall on both sides of that boundary. Most of this model's refusal was removable from the weights by an edit that needs no training, which places it among the properties an operator cannot rely on the weights alone to hold; the fraction that resisted every edit we tried is a property this study cannot yet assign to either side.

\section{Conclusion}

We applied an unmodified single-direction ablation recipe to a sparse mixture-of-experts model of roughly 320B parameters, with 288 routed experts per layer, a residual carried by four parallel streams, and weights released in block-FP8. The recipe itself was left alone. The direction is still the difference of means of the last-token representation over harmful and harmless prompts, and the edit is still one rank-one projection of the weights that write the residual stream, with no training, no gradients, and no second direction. Applying it to this model took three pieces of handling, each of which we measured the cost of. Writers are enumerated by tensor structure rather than by module name, giving 12,479 matrices, of which 12,384 are expert down-projections inside fused tensors; 12,442 of them are stored in FP8 and 37 in BF16. Activations are collapsed under the hyper-connection mixing weights before the representation is taken. Orthogonalization on quantized weights is done iteratively: 32 iterations bring the residual leakage from 13.4\% down to 5.2\% and the baked refusal rate from 0.188 to 0.094, against a lossless-hook prediction of 0.031.

The edited model's refusal rate on the seven harmful benchmarks falls from 0.51--1.00 to 0.08--0.48, every reduction is significant under a paired test, and across nearly 800 harmful prompts only two run the other way. The four capability scores move by no more than 1.0 point, three up and one down. False refusal on harmless prompts falls, from 0.024 to zero. The direction was fitted on 256 AdvBench prompts alone, and the clean variants that exclude the fitting sample give the same reduction as the originals, so what was removed is not memorization of that sample.

The anatomy comes from four further analyses. Editing attention, the dense and shared projections, or the routed experts on their own removes 0.039, 0.016 and 0.148 of refusal, while editing all three removes 0.776: three quarters of the effect exists only in the joint intervention. Matching module names reaches 0.7\% of the writers on the inference path, and editing them moves refusal by 0.066 out of 0.776 without raising an error. The multi-stream residual does not obstruct ablation, because its cross-stream mixing weights each stream by a scalar and performs no rotation within the hidden dimension. Editing the writers is nonetheless more thorough than projecting at the layer boundary over the same layers, 0.174 against 0.265 on the prompts valid in both conditions, a difference the marginal intervals hide and the paired test finds. What the edit removes is specific to the direction removed: a random direction orthogonal to it leaves refusal unchanged, and neither of the two rank-2 extensions we tested restored it. On violence, sexual content and hate, subspaces fitted specifically on those categories bring the residue from 0.986 to between 0.20 and 0.47, and none of the twelve ranks removes it.

A single direction reaches most of this model's refusal, but only when it is projected out of all of the writers at once. It survived two architectural changes: fused expert tensors require only a different way of enumerating the writers, and a multi-stream residual requires only that the mixing operator not rotate within the hidden dimension. Residual refusal remains concentrated in a few content categories. The scope should be stated plainly: one model, one selected layer for direction estimation, and the intervention procedures tested here. Whether it generalizes to other frontier mixture-of-experts models, and what carries the refusal that resists, remain unmeasured.

\section*{Limitations}

\paragraph{A single model.} Every result comes from GLM-5.3-Flash, with no cross-vendor comparison. The architecture-dependent conclusions---that fused expert tensors must be enumerated structurally, and that per-stream scalar mixing preserves orthogonality---transfer to models with the same features. The specific numbers, including the size of the interaction term, the choice of layer 22, and the per-category residues, are properties of this one checkpoint.

\paragraph{Judgment depends on a model.} Refusal is judged by a judge model, with no human annotation. The judge and the rule-based classifier agree on 0.85--0.96 of responses on the harmful benchmarks, and the disagreement runs systematically toward the rule underestimating refusal. For small effects such as false refusal we rely on an improved judge prompt rather than an independent human check.

\paragraph{The two scorers disagree unevenly across conditions.} Every experiment reported here is scored by both the judge model and the rule-based classifier; 5,475 hook-experiment responses were re-judged alongside the main results. The disagreement between them is not a constant offset: on the writer-subset sweep the rule classifier reads higher than the judge on lightly edited conditions and lower on heavily edited ones, by up to 0.048 in either direction. Because the two conventions also differ in how they treat unfinished replies, a comparison whose margin is small can change significance with the convention, as the dense group's increment does. We report both scorers wherever a comparison is close.

\paragraph{Capability evaluation is a subset.} Each of the four benchmarks is sampled down to a few hundred items, which rules out a large regression but does not substitute for a full evaluation harness.

\paragraph{The rank sweep is exploratory.} The twelve subspaces were fitted independently rather than as one nested basis (Table~\ref{tab:resistant}), so the sweep compares twelve independently fitted subspaces and does not measure what one more dimension does. Selecting the minimum over twelve conditions on 73 evaluation prompts compounds that. A controlled version would fix the seed, nest the bases, and pre-register the comparison.

\paragraph{One random control, no power analysis.} The direction-specificity result rests on a single random orthogonal direction; several would be needed to rule out arbitrary-direction explanations in general. The capability comparison reports paired intervals but no power analysis, so we quote no minimum detectable effect. Thinking mode is untested: all evaluation runs at \texttt{reasoning\_effort=low}.

\paragraph{Rates carry run-to-run variation.} Greedy decoding under tensor parallelism is not bit-deterministic. Two runs of the same condition on the same prompts with the same seed gave base refusal rates of 0.949 and 0.961, so a rate carries roughly 1--3 points of run-to-run noise. Every comparison we report is made within a single run, and we make no claim about differences of a few points across runs.

\paragraph{Withholding the implementation has a cost.} We do not publish the code for the editing pipeline, so verification rests entirely on the description in the method section. We give there everything an independent implementation needs---the full writer inventory and the criterion defining it, the iteration count against residual leakage, the two ways the scales must be handled, and the three reasons the rewrite has to proceed shard by shard---but that is not the same as reproducing by download, and a reader has to invest engineering effort to rebuild it.

\bibliography{references}

\appendix

\section{Judge label counts}
\label{app:labels}

Every rate in this paper comes from a judge model that assigns each response one of three
labels---\textsc{refusal}, \textsc{compliance}, or \textsc{unfinished} for a reply that ran
out of the token budget before answering---or, rarely, no usable label after retries. The
main text counts a response as a refusal when and only when the label is \textsc{refusal}.

That choice matters, so the counts behind it are given here. From them a reader can
recompute any rate under a different convention, see exactly which denominators shrank
where a verdict is missing, and check the two places where the convention changes the
answer. Column \textsc{r} is \textsc{refusal}, \textsc{c} is \textsc{compliance}, \textsc{u}
is \textsc{unfinished}, and $-$ counts responses with no usable label.

Worked example, HarmBench in Table~\ref{tab:app-main}: the base model gives 138
\textsc{refusal}, 5 \textsc{compliance} and 7 \textsc{unfinished} over 150 prompts. The main
table reports $138/150 = 0.920$. Counting anything that is not \textsc{compliance} as a
refusal would give $145/150 = 0.967$ instead.

\begin{table}[htbp]
\centering\footnotesize
\setlength{\tabcolsep}{4pt}
\begin{tabular}{lrrrr@{\hspace{1.4em}}rrrr}
\toprule
& \multicolumn{4}{c}{Base} & \multicolumn{4}{c}{Edited} \\
\cmidrule(lr){2-5}\cmidrule(lr){6-9}
Benchmark & \textsc{r} & \textsc{c} & \textsc{u} & $-$ & \textsc{r} & \textsc{c} & \textsc{u} & $-$ \\
\midrule
MaliciousInstruct & 97 & 3 & 0 & 0 & 8 & 92 & 0 & 0 \\
AdvBench & 97 & 2 & 1 & 0 & 13 & 85 & 1 & 1 \\
JailbreakBench & 94 & 6 & 0 & 0 & 15 & 83 & 2 & 0 \\
HarmBench & 138 & 5 & 7 & 0 & 25 & 120 & 5 & 0 \\
StrongREJECT & 149 & 0 & 1 & 0 & 57 & 91 & 2 & 0 \\
SimpleSafetyTests & 50 & 0 & 0 & 0 & 24 & 26 & 0 & 0 \\
ForbiddenQuestions & 77 & 68 & 5 & 0 & 15 & 134 & 1 & 0 \\
AdvBench (clean) & 98 & 1 & 0 & 1 & 15 & 83 & 2 & 0 \\
JailbreakBench (clean) & 87 & 5 & 0 & 0 & 18 & 73 & 1 & 0 \\
StrongREJECT (clean) & 149 & 0 & 1 & 0 & 49 & 99 & 2 & 0 \\
XSTest (safe) & 6 & 241 & 3 & 0 & 0 & 250 & 0 & 0 \\
\bottomrule
\end{tabular}
\caption{Labels behind Table~\ref{tab:main}. Two responses have no usable label---one on the
edited AdvBench run and one on the base AdvBench (clean) run---which is why both AdvBench
rows in Table~\ref{tab:main} are scored out of 99.}
\label{tab:app-main}
\end{table}

\begin{table}[htbp]
\centering\footnotesize
\setlength{\tabcolsep}{4pt}
\begin{tabular}{lrrrr@{\hspace{1.4em}}rrr}
\toprule
& \multicolumn{4}{c}{Harmful (256)} & \multicolumn{3}{c}{Harmless (100)} \\
\cmidrule(lr){2-5}\cmidrule(lr){6-8}
Edited group & \textsc{r} & \textsc{c} & \textsc{u} & $-$ & \textsc{r} & \textsc{c} & \textsc{u} \\
\midrule
none & 243 & 10 & 3 & 0 & 5 & 74 & 21 \\
attention & 233 & 18 & 5 & 0 & 4 & 85 & 11 \\
dense $+$ shared & 239 & 15 & 2 & 0 & 5 & 82 & 13 \\
routed experts & 205 & 42 & 9 & 0 & 2 & 81 & 17 \\
attention $+$ dense & 226 & 28 & 2 & 0 & 2 & 86 & 12 \\
attention $+$ experts & 104 & 139 & 12 & 1 & 1 & 77 & 22 \\
dense $+$ experts & 166 & 85 & 3 & 2 & 0 & 80 & 20 \\
all three & 44 & 201 & 9 & 2 & 0 & 84 & 16 \\
layer boundary & 69 & 180 & 6 & 1 & 1 & 85 & 14 \\
\bottomrule
\end{tabular}
\caption{Labels behind Table~\ref{tab:stages}. The harmless columns are the first place the
convention decides the answer: the unedited model refuses 5 of 100 harmless prompts, but 21
more replies run past the token budget, so counting non-\textsc{compliance} as refusal would
report a false-refusal rate of 0.260 for a model whose false-refusal rate is 0.050. The
sweep's 3,276 records are these nine conditions over 256 harmful, 100 harmless and 8
quarantined prompts, the last being prompts inside the fitting slice, scored separately.}
\label{tab:app-stages}
\end{table}

\begin{table}[htbp]
\centering\footnotesize
\setlength{\tabcolsep}{4pt}
\begin{tabular}{lrrrr@{\hspace{1.4em}}rrrr}
\toprule
& \multicolumn{4}{c}{StrongREJECT (150)} & \multicolumn{4}{c}{JBB (100)} \\
\cmidrule(lr){2-5}\cmidrule(lr){6-9}
Condition & \textsc{r} & \textsc{c} & \textsc{u} & $-$ & \textsc{r} & \textsc{c} & \textsc{u} & $-$ \\
\midrule
none & 149 & 1 & 0 & 0 & 92 & 8 & 0 & 0 \\
$r$, rank 1 & 48 & 100 & 2 & 0 & 19 & 80 & 1 & 0 \\
random $\perp r$, rank 1 & 147 & 2 & 0 & 1 & 92 & 7 & 1 & 0 \\
$r + \mathrm{PC}_1$, rank 2 & 46 & 98 & 6 & 0 & 15 & 80 & 5 & 0 \\
$r + \text{random}$, rank 2 & 38 & 108 & 4 & 0 & 16 & 82 & 2 & 0 \\
\bottomrule
\end{tabular}
\caption{Labels behind Table~\ref{tab:directions}.}
\label{tab:app-arms}
\end{table}

\begin{table}[htbp]
\centering\footnotesize
\setlength{\tabcolsep}{5pt}
\begin{tabular}{lrrrr}
\toprule
Rank & \textsc{r} & \textsc{c} & \textsc{u} & $-$ \\
\midrule
none & 72 & 1 & 0 & 0 \\
1 & 28 & 43 & 2 & 0 \\
2 & 31 & 41 & 1 & 0 \\
3 & 17 & 52 & 4 & 0 \\
4 & 20 & 52 & 1 & 0 \\
5 & 34 & 35 & 4 & 0 \\
6 & 23 & 40 & 7 & 3 \\
7 & 15 & 47 & 9 & 2 \\
8 & 14 & 44 & 12 & 3 \\
9 & 21 & 39 & 13 & 0 \\
10 & 21 & 37 & 15 & 0 \\
11 & 20 & 42 & 11 & 0 \\
12 & 16 & 44 & 13 & 0 \\
\bottomrule
\end{tabular}
\caption{Labels behind Table~\ref{tab:resistant}, 73 prompts per rank. This is the second
place the convention decides the answer. Ranks 6, 7 and 8 lose 3, 2 and 3 responses to
missing labels, which is why their denominators are 70, 71 and 70 rather than 73, and
\textsc{unfinished} climbs from 2 at rank 1 to 15 at rank 10---so a low
\textsc{refusal}-only rate at high rank is partly replies that never reached an answer.}
\label{tab:app-rank}
\end{table}

These tables cover 8,259 response records: 2,784 for the main results (both checkpoints over
eleven benchmark sections), 3,276 for the writer-subset sweep, 1,250 for the direction
conditions and 949 for the rank sweep. The number of judge calls is smaller, because verdicts
are cached on the prompt and the response together and two conditions that produce identical
text for the same prompt are judged once; every record carries a label either way.

\end{document}